\documentclass[12pt,epsf,a4paper]{article}
\usepackage{graphicx,amsmath,amssymb}
\usepackage{cite}
\usepackage{amsfonts}
\usepackage[height=22.5cm,width=16.5cm]{geometry}
\begin{document}
%%%%%%%%%%%%%%%%%%%%%%%%%%%%%%%%%%%%%%%%%%%

\def\a{\alpha}
\def\b{\beta}
\def\c{\varepsilon}
\def\d{\delta}
\def\e{\epsilon}
\def\f{\phi}
\def\g{\gamma}
\def\h{\theta}
\def\k{\kappa}
\def\l{\lambda}
\def\m{\mu}
\def\n{\nu}
\def\p{\psi}
\def\q{\partial}
\def\r{\rho}
\def\s{\sigma}
\def\t{\tau}
\def\u{\upsilon}
\def\v{\varphi}
\def\w{\omega}
\def\x{\xi}
\def\y{\eta}
\def\z{\zeta}
\def\D{\Delta}
\def\G{\Gamma}
\def\H{\Theta}
\def\L{\Lambda}
\def\F{\Phi}
\def\P{\Psi}
\def\S{\Sigma}

\def\o{\over}
\def\beq{\begin{eqnarray}}
\def\eeq{\end{eqnarray}}
\newcommand{\gsim}{ \mathop{}_{\textstyle \sim}^{\textstyle >} }
\newcommand{\lsim}{ \mathop{}_{\textstyle \sim}^{\textstyle <} }
\newcommand{\vev}[1]{ \left\langle {#1} \right\rangle }
\newcommand{\bra}[1]{ \langle {#1} | }
\newcommand{\ket}[1]{ | {#1} \rangle }
\newcommand{\EV}{ {\rm eV} }
\newcommand{\KEV}{ {\rm keV} }
\newcommand{\MEV}{ {\rm MeV} }
\newcommand{\GEV}{ {\rm GeV} }
\newcommand{\TEV}{ {\rm TeV} }
\def\diag{\mathop{\rm diag}\nolimits}
\def\Spin{\mathop{\rm Spin}}
\def\SO{\mathop{\rm SO}}
\def\O{\mathop{\rm O}}
\def\SU{\mathop{\rm SU}}
\def\U{\mathop{\rm U}}
\def\Sp{\mathop{\rm Sp}}
\def\SL{\mathop{\rm SL}}
\def\tr{\mathop{\rm tr}}

\def\IJMP{Int.~J.~Mod.~Phys. }
\def\MPL{Mod.~Phys.~Lett. }
\def\NP{Nucl.~Phys. }
\def\PL{Phys.~Lett. }
\def\PR{Phys.~Rev. }
\def\PRL{Phys.~Rev.~Lett. }
\def\PTP{Prog.~Theor.~Phys. }
\def\ZP{Z.~Phys. }

%%%%%%% added by Fumi %%%%%%%%%%
% FROM HERE
%\newcommand{\beq}{\begin{equation}}   
%\newcommand{\eeq}{\end{equation}}
\newcommand{\bea}{\begin{eqnarray}}   
\newcommand{\eea}{\end{eqnarray}}
\newcommand{\bear}{\begin{array}}  
\newcommand {\eear}{\end{array}}
\newcommand{\bef}{\begin{figure}}  
\newcommand {\eef}{\end{figure}}
\newcommand{\bec}{\begin{center}}  
\newcommand {\eec}{\end{center}}
\newcommand{\non}{\nonumber}  
\newcommand {\eqn}[1]{\beq {#1}\eeq}
\newcommand{\la}{\left\langle}  
\newcommand{\ra}{\right\rangle}
\newcommand{\ds}{\displaystyle}
\def\SEC#1{Sec.~\ref{#1}}
\def\FIG#1{Fig.~\ref{#1}}
\def\EQ#1{Eq.~(\ref{#1})}
\def\EQS#1{Eqs.~(\ref{#1})}
\def\GEV#1{10^{#1}{\rm\,GeV}}
\def\MEV#1{10^{#1}{\rm\,MeV}}
\def\KEV#1{10^{#1}{\rm\,keV}}
\def\lrf#1#2{ \left(\frac{#1}{#2}\right)}
\def\lrfp#1#2#3{ \left(\frac{#1}{#2} \right)^{#3}}
% UNTIL HERE

%%%%%%%%%%%%%%%%%%%%%%%%%%%%%%%%%%%%%%%%%%%%%%%%%%%%%%%%%%%%%%%%%%%%

%\baselineskip 0.7cm

\begin{titlepage}

\begin{flushright}
UT-13-36
\end{flushright}

\vskip 3cm
\begin{center}
{\large \bf 
	The universe dominated by oscillating scalar with\\[.2em] non-minimal derivative coupling to gravity
}
\vskip 1.2cm
Ryusuke Jinno$^a$,
Kyohei Mukaida$^a$ 
and
Kazunori Nakayama$^{a,b}$

\vskip 0.7cm

\begin{tabular}{ll}
$^a$ \!\!\!\!\!\!& {\em Department of Physics, Faculty of Science, }\\
& {\em University of Tokyo,  Bunkyo-ku, Tokyo 133-0033, Japan}\\[.5em]
$^b$ \!\!\!\!\!\!& {\em Kavli Institute for the Physics and Mathematics of the Universe (WPI), }\\
&{\em Todai Institute for Advanced Study, }\\
&{\em University of Tokyo,  Kashiwa, Chiba 277-8583, Japan}
\end{tabular}

\vskip 1.5cm

\abstract{
	We study the expansion law of the universe dominated by the oscillating scalar field with non-minimal derivative
	coupling to gravity as $G^{\mu \nu} \partial_{\mu} \phi \partial_{\nu} \phi$. 
	In this system the Hubble parameter oscillates with a frequency of the effective mass of the scalar field, 
	which formerly caused a difficulty in analyzing how the universe expands. 
	We find an analytical solution for power law potentials and interpret the solution in an intuitive way 
	by using a new invariant of the system. As a result, we find marginally accelerated expansion for the quadratic potential and 
	no accelerated expansion for the potential with higher power. 
	}
\end{center}
\end{titlepage}

\setcounter{page}{2}

%%%%%%%%%%%%%%%%%%%%%%%%%%%%%%%%%%%%%
\section{Introduction}
%%%%%%%%%%%%%%%%%%%%%%%%%%%%%%%%%%%%%

After the discovery of the standard model (SM) like Higgs boson at the LHC~\cite{Aad:2012tfa}
and the results from the Planck satellite~\cite{Ade:2013uln}, one class of the well-motivated inflation models is the Higgs inflation models.
The Higgs inflation models utilize a non-minimal coupling of the inflaton $\phi$ to gravity and/or non-minimal kinetic term.\footnote{
	Although these non-minimal inflation models are motivated by the possibility of the SM Higgs
	boson as the inflaton $\phi$, the same inflationary dynamics is caused by some other scalar fields
	with similar non-minimal couplings. Thus we do not necessarily identify $\phi$ as the SM Higgs boson hereafter.
}
There are several types of the Higgs inflation proposed so far:
coupling with the Ricci scalar $\sim R \phi^2$~\cite{Bezrukov:2007ep},
coupling to the Einstein tensor $\sim G^{\mu\nu}\partial_\mu\phi\partial_\nu\phi$~\cite{Germani:2010gm}
and the non-minimal kinetic term $\sim \phi^n (\partial\phi)^2$~\cite{Nakayama:2010kt}.
See Refs.~\cite{Kobayashi:2011nu,Kamada:2012se} for more general class of models with non-minimal couplings without introducing extra degrees of freedom.
Constraints on these models in light of the Planck results are found in Ref.~\cite{Tsujikawa:2013ila}.

In order to predict the scalar spectral index of the density perturbation with a good accuracy, we need to know the details of the reheating process.
In the case of minimal canonical kinetic term for the oscillating inflaton, the expansion law for the power-law potential is well-known~\cite{Turner:1983he}.
In non-minimal models such as the Higgs inflation, however, the expansion law during the inflaton oscillation era may be rather complicated due to the modified equation of motion of the scalar field.
In particular, models with non-minimal derivative couplings to gravity $(\sim G^{\mu\nu}\partial_\mu\phi\partial_\nu\phi) $
lead to an unusual equation of motion of the scalar field~\cite{Germani:2010gm} 
(see also Refs.~\cite{Amendola:1993uh,Capozziello:1999xt,Granda:2009fh})
and deriving the expansion law in the universe dominated by such a scalar is non-trivial.

In this paper we study the evolution of the oscillating scalar with a non-minimal coupling to gravity 
$(\sim G^{\mu\nu}\partial_\mu\phi\partial_\nu\phi)$ in detail and derive the expansion law of the universe.
In the main text, we first show that the energy density of $\phi$ is not a good conserved quantity in a time scale of $\phi$ oscillation.
Instead, we find another useful invariant, which we will call $J$. 
Using this, we will derive the expansion law of the universe in a rather simple way. 
Moreover, we find an analytical solution for the scalar dynamics when the non-minimal kinetic term plays a dominant role.
The analytic solution is illustrated in detail in Appendix and it explicitly exhibits the existence of the invariant $J$.
It is fully consistent with the results in the main text obtained in a more intuitive way.\footnote{
	Our results are inconsistent with previous studies~\cite{Sadjadi:2012zp,Ghalee:2013ada,Sadjadi:2013psa}.
}

%%%%%%%%%%%%%%%%%%%%%%%%%%%%%%%%%%%%%
\section{Analysis}   \label{sec:ana}
%%%%%%%%%%%%%%%%%%%%%%%%%%%%%%%%%%%%%

We consider the following action with a real scalar field $\phi$, 
\begin{equation}
	S = \int d^4 x \sqrt{-g} \left[\frac{M_P^2}{2}R -\frac{1}{2}\left(g^{\mu\nu} -\frac{G^{\mu\nu}}{M^2}\right)\partial_\mu\phi\partial_\nu\phi 
	-V(\phi) \right].   \label{S}
\end{equation}
where $G^{\mu\nu} = R^{\mu\nu}-Rg^{\mu\nu}/2$ is the Einstein tensor, $M_P$ the reduced Planck scale, $R$  the Ricci scalar
and the Friedmann-Robertson-Walker metric is defined by
\begin{equation}
	ds^2 = g_{\mu\nu}dx^\mu dx^\nu = - dt^2 +a^2(t)(dx^2+dy^2+dz^2),
\end{equation}
with $a(t)$ being the scale factor.
By the standard procedure, we find the energy density and pressure of the scalar field as
\begin{equation}
	\rho_\phi = \left(1+\frac{9H^2}{M^2}\right)\frac{\dot\phi^2}{2} + V,
	\label{rho}
\end{equation}
and
\begin{equation}
	p_\phi = \left(1-\frac{3H^2}{M^2} \right)\frac{\dot\phi^2}{2}-V
	-\frac{1}{M^2} \frac{d}{dt}(H\dot\phi^2),
	\label{p}
\end{equation}
where $H=\dot a/a$ is the Hubble parameter.
They satisfy the following energy conservation ensured by the Bianchi identity:
\begin{equation}
	\dot \rho_\phi + 3H (\rho_\phi + p_\phi) = 0.
	\label{dotrho}
\end{equation}
Assuming that the scalar field dominates the universe, the Friedmann equation reads
\begin{equation}
	H^2 = \frac{\rho_\phi}{3M_P^2}  ~~~\leftrightarrow~~~ 
	H^2 = \frac{1}{3M_P^2}\frac{\frac{1}{2}\dot\phi^2+V}{1-\frac{3\dot\phi^2}{2M^2M_P^2}}.
	\label{Fried}
\end{equation}
On the other hand, the equation of motion of the scalar field is given by
\begin{equation}
	\left( 1+ \frac{3H^2}{M^2} \right)\ddot\phi + 3H\left(1+ \frac{3H^2}{M^2}+\frac{2\dot H}{M^2} \right)\dot\phi + 
	\frac{\partial V}{\partial \phi} = 0.
	\label{eqofm_full}
\end{equation}
Here $\dot H$ is calculated from (\ref{Fried}) and (\ref{eqofm_full}) as
\begin{equation}
	\frac{\dot H}{H} = -\frac{\left(1+\frac{9H^2}{M^2} \right)3H\dot\phi^2 + \frac{6H^2}{M^2+3H^2}\frac{\partial V}{\partial \phi}\dot\phi  }
	{\left[\left(1+\frac{9H^2}{M^2} \right)\frac{\dot\phi^2}{2} + V \right] \left[ 2-\left(1-\frac{9H^2}{M^2} \right)\frac{\dot\phi^2}{M_P^2(M^2+3H^2)}  \right] }.
	\label{dotH_full}
\end{equation}
It is soon realized that the equation of motion (\ref{eqofm_full}) reduces to the standard one for small $H$ limit.\footnote{
	Terms including $M$ are neglected for $H \ll M(M/m_{\rm eff})^{1/3}$ where $m_{\rm eff} \sim \sqrt{V_{,\phi}/\phi}$
	is the effective inflaton mass ($V_{,\phi}\equiv \partial V/\partial \phi$).
}
We are interested in the opposite case $H \gg M$, where the non-minimal kinetic term takes an important role,
hence we consider this case in the following.\footnote{
	In such a case, higher dimensional operators may also become important.
	Since we do not know UV complete gravity theory, we simply assume that the term in the action (\ref{S}) is dominant.
}

So let us take the limit $H \gg M$.
Then the equation of motion reduces to
\begin{equation}
	\ddot\phi + 3H\left(1+ \frac{2\dot H}{3H^2} \right)\dot\phi + 
	\frac{M^2}{3H^2}\frac{\partial V}{\partial \phi} = 0.
	\label{eqofm}
\end{equation}
In this limit, Eq.~(\ref{dotH_full}) becomes
\begin{equation}
	\frac{\dot H}{H} = -\frac{\frac{27H^3}{2M^2}\dot\phi^2 + \frac{\partial V}{\partial \phi} \dot\phi }{\frac{9H^2}{M^2}\dot\phi^2 + V}.
	\label{dotH}
\end{equation}
Hereafter we assume the power-law potential 
\begin{equation}
	V = \frac{\lambda}{n}\phi^n.
	\label{V}
\end{equation}
We have $n=4$ in the case of SM Higgs boson.
The equations to be solved are Eq.~(\ref{eqofm}) with Eqs.~(\ref{Fried}) and (\ref{dotH})
with initial conditions, say, $\phi=\phi_i$ and $\dot \phi = 0$.

There are several remarks on this system.
First, it is seen from Eq.~(\ref{eqofm}) that $\phi$ undergoes a coherent oscillation around the potential
with a frequency $m_{\rm eff} \sim (M/H)\sqrt{V_{,\phi}/\phi}$ for $m_{\rm eff} \gg H$.
In the following, we must be careful on the distinction between ``fast'' variables which oscillate with frequency $\sim m_{\rm eff}$
and ``slow'' variables which only change with the expansion rate $(\sim H)$. 
We take a limit $H/m_{\rm eff} \ll 1$ hereafter, since otherwise inflation might happen.
What is unusual in the present model is that $H$ itself is not a slow variable, but a fast variable:
$\dot H \sim m_{\rm eff} H$, not $\dot H \sim H^2$.
To see this, note that $\rho_\phi$ and $p_\phi$ are not the same order: the last term in Eq.~(\ref{p}) makes a large contribution
and hence $\rho_\phi \sim \mathcal O(m_{\rm eff}^2\phi_0^2)$ and $p_\phi \sim \mathcal O(m_{\rm eff}^3\phi_0^2 H^{-1})$
where $\phi_0$ is the amplitude of the oscillating scalar field.
Thus the last term in the energy conservation (\ref{dotrho}), $\sim Hp_\phi$, is fairly large
and we obtain $\dot\rho_\phi \sim \mathcal O(m_{\rm eff} \rho_\phi)$.
It means that $\rho_\phi$ is not a conserved quantity in an oscillation time scale.
Rather, $\rho_\phi$ oscillates with the time scale of $m_{\rm eff}$, so does $H$.
This fact makes the analysis complicated compared with the standard case with canonical kinetic term.

Fortunately, we find a good conserved quantity:\footnote{
	This invariant will be explicitly derived in Appendix with an analytical solution.
}
\begin{equation}
	J \equiv \frac{1}{H}\left( \frac{3H^2}{M^2}\dot\phi^2 + V \right).
	\label{J}
\end{equation}
Actually, using Eqs.~(\ref{eqofm}) and (\ref{dotH}), we obtain
\begin{equation}
	\dot J = -\frac{9}{2}\frac{H^2}{M^2}\dot\phi^2.
	\label{dotJ}
\end{equation}
Note that the RHS is $\sim \mathcal O(H J)$.
Thus $J$ is a slow variable which only changes due to the Hubble expansion.
Since it is conserved in an oscillation of the scalar field, we can evaluate it at $\phi = \phi_0(t)$ and $\dot\phi=0$:
\begin{equation}
	J = \sqrt{\frac{3\lambda}{n}}M_P\phi_0(t)^{n/2}.
\end{equation}
Here $\phi_0(t)$ is the amplitude of the scalar field oscillation, which also slowly varies with time due to the Hubble expansion.
By taking time average, Eq.~(\ref{dotJ}) becomes
\begin{equation}
	\dot J = -\frac{9}{2M^2} \langle H^2 \dot\phi^2 \rangle.
\end{equation}
Here $\langle\cdots\rangle$ means the time average with respect to $t$ over the oscillation time scale $\sim m_{\rm eff}^{-1}$
which is much shorter than the Hubble time scale $\sim H^{-1}$.
Within this time scale, slow variables are regarded as constants.
Now let us define the following time averaged quantities,
\begin{gather}
	\langle \phi^n \rangle \equiv c \phi_0^n,  \label{c} \\
	\langle K \rangle \equiv d \langle V \rangle,\label{d}  \\
	\langle H^2 \rangle \equiv e \langle H \rangle ^2, \label{e} 
\end{gather}
where $c,d,e$ are numerical constants of order unity and $K\equiv (9H^2/2M^2)\dot\phi^2$.
In the present case, we find $e= 9(1+d)/[c(3+2d)^2]$.
Then we find
\begin{equation}
	\phi_0(t)^{n/2} = \left[\frac{1}{\phi_0(t_0)^{n/2}} + cd\sqrt{\frac{\lambda}{3n}}\frac{1}{M_P}(t-t_0)  \right]^{-1},
\end{equation}
for the initial condition $\phi_0 = \phi_0(t_0)$ at $t=t_0$.
Thus it scales as $\phi_0(t) \propto t^{-2/n}$ and $\langle\rho_\phi\rangle \propto t^{-2}$.
Using this, we obtain
\begin{equation}
	\langle H\rangle \simeq \frac{3+2d}{3d}\frac{1}{t} \equiv \frac{\epsilon}{t}~~~
	{\rm for}~~~t \gtrsim \sqrt{\frac{3n}{\lambda}}\frac{M_P}{\phi_0(t_0)^{n/2}}\sim H(t_0)^{-1}.
	\label{explaw}
\end{equation}
This is the expansion law that we have been looking for.
The scale factor varies with time\footnote{
	Precisely speaking, $\langle H \rangle \neq \dot{\langle a\rangle}/\langle a \rangle$.
	But the difference is not relevant for the present purpose.
}
as $\langle a\rangle \propto t^\epsilon$.
The remaining task is to derive numerical constants $c,d,e$.
This is done by numerically solving the equation of motion (\ref{eqofm}).
In the present model, fortunately, there is a full analytic solution. See Appendix for detail.
The resulting constants are summarized in Table~\ref{tab:gamma} for $n=2$ and $n=4$.
For $n=2$, we obtain $\epsilon=1$, meaning the marginally accelerated expansion of the universe.
See Ref.~\cite{Damour:1997cb} for the phenomena known as the oscillating inflation.

\begin{table}[t]
\caption{Numerical constants.}
\begin{center}
  \begin{tabular}{c|cccc}
    \hline
     $n$ & $c$ & $d$ & $e$ & $\epsilon$ \\
    \hline
    $2$ & $0.438$& $3$ & $1.016$ & $1$ \\
    $4$ & $0.277$ & $6$ &  $1.012$ & $0.833$\\
    \hline
  \end{tabular}
\label{tab:gamma}
\end{center}
\end{table}

\begin{figure}
\begin{center}
\includegraphics[scale=1.2]{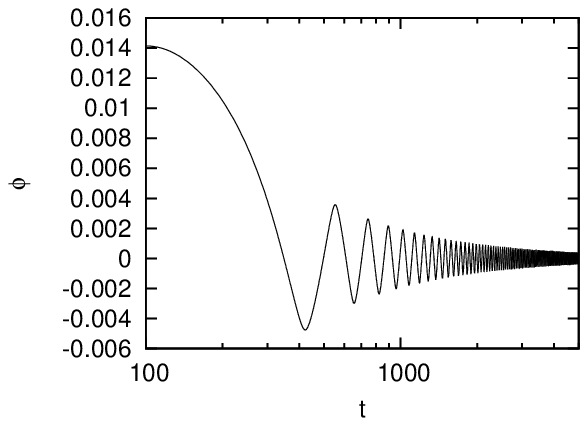}
\includegraphics[scale=1.2]{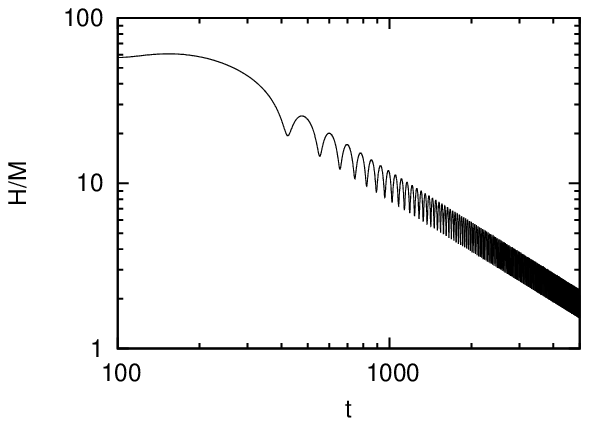}
\includegraphics[scale=1.2]{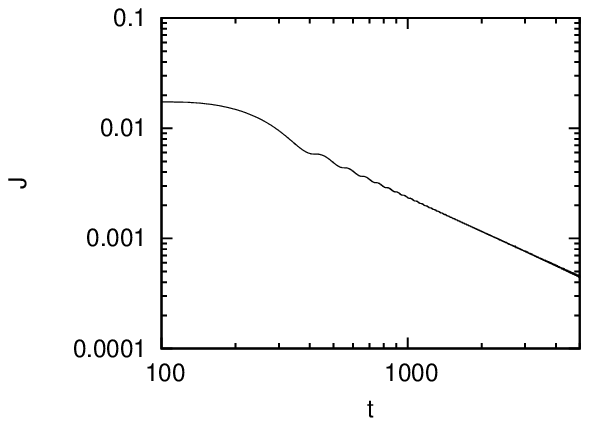}
\includegraphics[scale=1.2]{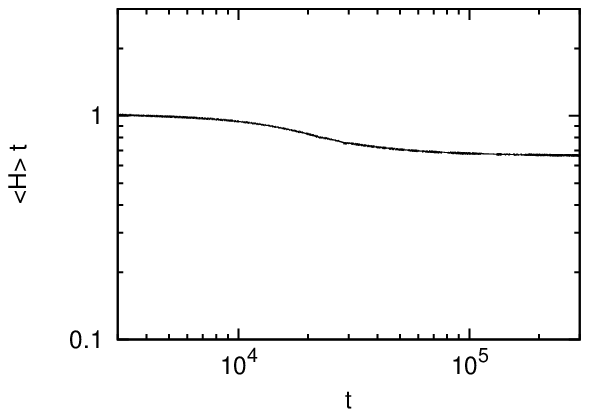}
\caption{ 
	Results of numerical calculation for $M=10^{-4}$, $\lambda=1$ and $n=2$ in the Planck unit.
	(Upper left) Time evolution of $\phi$.
	(Upper right) Time evolution of $H/M$.
	(Lower left) Time evolution of $J$.
	(Lower right) Time evolution of $\langle H\rangle t$.
}
\label{fig:n=2}
\end{center}
\end{figure}

\begin{figure}
\begin{center}
\includegraphics[scale=1.2]{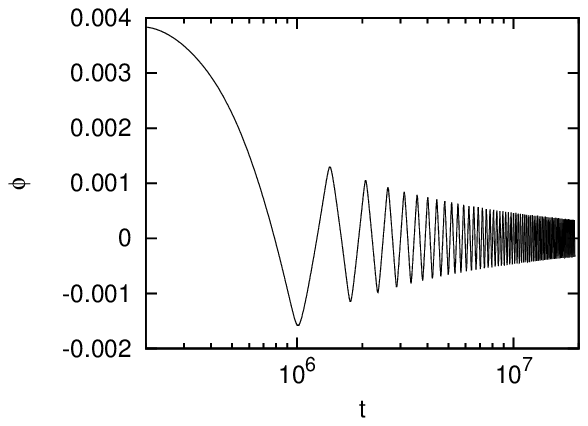}
\includegraphics[scale=1.2]{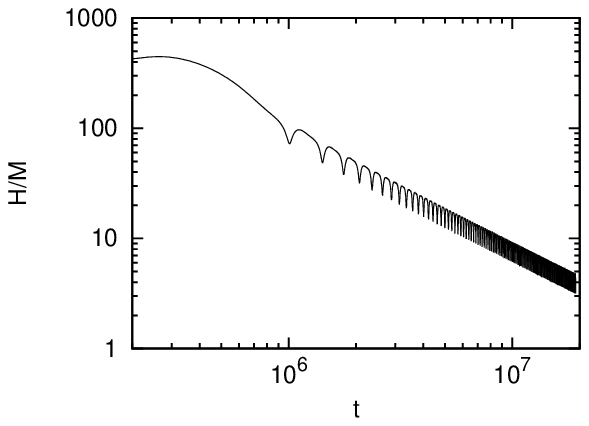}
\includegraphics[scale=1.2]{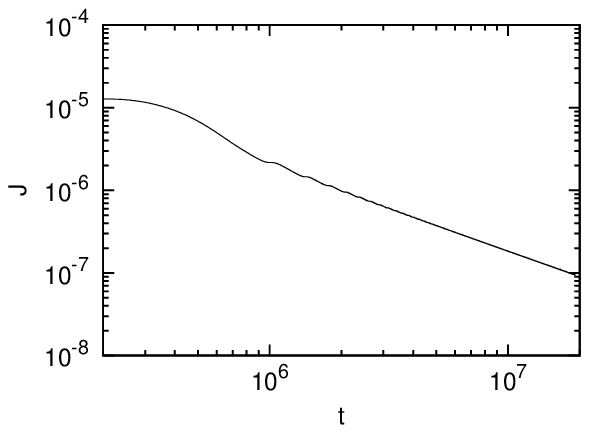}
\includegraphics[scale=1.2]{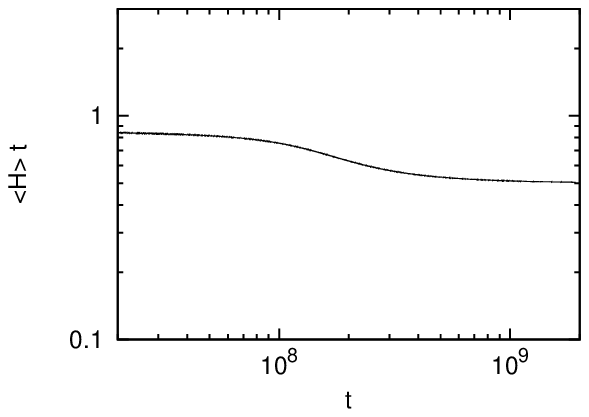}
\caption{ 
	Results of numerical calculation for $M=10^{-8}$, $\lambda=1$ and $n=4$ in the Planck unit.
	(Upper left) Time evolution of $\phi$.
	(Upper right) Time evolution of $H/M$.
	(Lower left) Time evolution of $J$.
	(Lower right) Time evolution of $\langle H\rangle t$.
}
\label{fig:n=4}
\end{center}
\end{figure}

We have performed numerical calculation to check the above considerations.
We have solved Eq.~(\ref{eqofm_full}) with (\ref{Fried}).
The results are shown in Figs.~\ref{fig:n=2} and \ref{fig:n=4}.
Parameters are chosen as $M=10^{-4}$, $\lambda=1$, $n=2$ in the Planck unit in Fig.~\ref{fig:n=2},
and $M=10^{-8}$, $\lambda=1$, $n=4$ in Fig.~\ref{fig:n=4}.
The initial condition is set to be $\phi = (n^3 M^2 M_P^4/2\lambda)^{1/(n+2)}$ and $\dot\phi = 0$ where $m_{\rm eff}/H = 1$.
It is seen that the Hubble parameter oscillates with (twice) the frequency of $\phi$ oscillation, while $J$ does not, and the amplitude of the $H$ oscillation relative to the average value of $H$ remains constant.\footnote{
	Even in the case of a scalar with minimal canonical kinetic term, $H$ oscillates at the very beginning
	of the scalar oscillation. 
	However, the relative amplitude of $H$ soon damps and $H$ eventually becomes a 
	conserved quantity which only changes with the Hubble expansion.
}
It is also seen that $\langle H\rangle t$ is equal to $1$ $(0.833)$ for $H \gg M$ and it finally becomes $2/3$ $(1/2)$
as is expected for the case of the scalar oscillation with minimal canonical kinetic term in the quadratic (quartic) potential for $n=2$ $(n=4)$.
We have also checked that analytical results listed in Table~\ref{tab:gamma} are reproduced numerically.

%%%%%%%%%%%%%%%%
%%%%%%%%%%%%%%%%

%%%%%%%%%%%%%%%%
%%%%%%%%%%%%%%%%

%%%%%%%%%%%%%%%%%%%%%%%%%%%%%%%%%%%%%
\section{Summary and Discussion}
%%%%%%%%%%%%%%%%%%%%%%%%%%%%%%%%%%%%%

We have derived the expansion law of the universe dominated by the scalar field oscillation with the non-minimal derivative coupling to gravity,
especially when the non-minimal kinetic term dominates over the minimal canonical kinetic term.
Properties of the expansion law are significantly different from the standard picture:
the Hubble parameter oscillates with the same time scale as the scalar oscillation
and the (averaged) scale factor experiences an unusual expansion law.
This scalar oscillation dominated era is important since it necessarily appears after the inflation.
To summarize, the universe undergoes following phases in the time ordering:
\begin{itemize}
\item $m_{\rm eff} \ll H$ : inflation takes place.
\item $M\ll H \ll m_{\rm eff}$ : the universe is dominated by the oscillating scalar field with non-minimal kinetic term.
The expansion law is summarized in Table~\ref{tab:gamma}.
\item $M(M/m_{\rm eff})^{1/3} \ll H \ll M$ : the kinetic term is dominated by the canonical one,
but still the $\dot H$ term in the equation of motion is effective. 
The expansion law gradually approaches to the standard one, as shown in the lower right panel of Figs.~\ref{fig:n=2} and \ref{fig:n=4}.
\item $H\ll M(M/m_{\rm eff})^{1/3}$ :
the scalar dynamics is the standard one.
\end{itemize}
In a more realistic setup, the inflaton couples to the SM particles and it decays/dissipates into radiation.
The epoch at which the reheating is completed depends on the inflaton coupling to the SM fields.
Both the expansion law during the oscillating phase and the process of reheating
are crucial for predicting the scalar spectral index with high accuracy.
Since the Hubble parameter rapidly oscillates, it may have some indications on the dynamics of other scalar fields, inflationary gravitational waves and production of particles having Hubble masses.
Similar phenomena may also occur in models with other types of non-minimal kinetic terms.
We will return to these issues elsewhere.

%%%%%%%%%%%%%%%%%%%%%%%%%%%%%%%%%%%%%%%%%%%%
\section*{Acknowledgments}
%%%%%%%%%%%%%%%%%%%%%%%%%%%%%%%%%%%%%%%%%%%%

This work was supported by the Grant-in-Aid for Scientific Research on
Innovative Areas (No. 21111006  [K.N.]) and Scientific Research (A) (No. 22244030 [K.N.]).
The work of R.J. and K.M. is supported in part by JSPS Research Fellowships
for Young Scientists.

\appendix

%%%%%%%%%%%%%%%%%%%%%%%%%%%%%%%%%%%%%%%%%%%%
\section{Analytical solution} \label{sec:app}
%%%%%%%%%%%%%%%%%%%%%%%%%%%%%%%%%%%%%%%%%%%%

In this appendix we show that the equation of motion (\ref{eqofm}) has a very precise 
perturbative analytical solution in the oscillating regime. 
It is sufficiently precise already in the perturbation of the first order.
We will explicitly construct a conserved quantity introduced in the 
main text (Eq.(\ref{J})) by using the analytical solution. 
In addition we calculate the coefficients $c,d$, and $e$ in Sec.~\ref{sec:ana}, and 
show that the expansion law calculated in terms of these coefficients is consistent with the analytical solution.

%%%%%%%%%%%%%%%%%%%%%%%%%%%%%%%%%%%%%%%%%%%%%%%%%%
\subsection{Rewriting the equation}
%%%%%%%%%%%%%%%%%%%%%%%%%%%%%%%%%%%%%%%%%%%%%%%%%%

We start with Eq.~(\ref{eqofm}) with the power law potential (\ref{V}). 
We adopt the Planck unit $M_{P} = 1$. 
The initial value of $\phi$ is denoted by $\phi_{\rm ini}$ and we define
\begin{gather}
\phi
\equiv \phi_{\rm ini} \chi.
\end{gather}
Also we define the following dimensionless quantities:
\begin{gather}
h 
\equiv \frac{H}{M}, \\
\frac{d}{d\tau}
\equiv \frac{3h}{\mu M} \frac{d}{dt}, \\
\alpha
\equiv \left( \frac{9\lambda}{2nM^2} \right)^{1/2} \phi_{\rm ini}^{n/2+1} , \\
\mu
\equiv \left( \frac{2\lambda}{nM^2} \right)^{1/2} \phi_{\rm ini}^{n/2-1},
\end{gather}
Then we get
\begin{gather}
\chi'' + \left[ \frac{1}{2}(\ln(\chi^{'2}+\chi^n))' + \alpha (\chi^{'2} + \chi^n) \right]\chi' + \frac{3}{2}n\chi^{n-1} = 0,
\end{gather}
where the prime denotes the derivative with respect to $\tau$. 
Note that the term with $\alpha$ $(\sim H/m_{\rm eff})$ comes from the second derivative term and the conventional Hubble friction term in Eq.~(\ref{eqofm}). 
Since we are interested in the oscillating regime, we treat the $\alpha$-term as a small perturbation $(\alpha \ll 1)$. 
Next we define
\begin{gather}
\chi_1 
\equiv \chi^{n/2}, \\
\chi_2
\equiv \chi', \\
r
\equiv (\chi_1^2 + \chi_2^2)^{1/2}, \\
\tan \theta
\equiv \frac{\chi_2}{\chi_1},
\end{gather}
then we get
\begin{gather}
\frac{r'}{r} - \theta' \tan \theta - \frac{1}{1-\beta} (r \cos \theta)^{\beta} \tan \theta 
= 0, 
\label{eq_eom_rtheta_1} \\
2\frac{r'}{r}\tan \theta + \theta' + \alpha r^2 \tan \theta +\frac{3}{1-\beta} (r \cos \theta)^{\beta}
= 0,
\label{eq_eom_rtheta_2}
\end{gather}
where
\begin{gather}
\beta
\equiv -\frac{2}{n} + 1.
\end{gather}
These equations are rewritten as
\begin{gather}
r'
= -\frac{1}{1-\beta} r^{1+\beta} \frac{t_{\theta}}{(t_{\theta}^2+1/2)(t_{\theta}^2+1)^{\beta/2}} - \frac{\alpha}{2} r^3 \frac{t_{\theta}^2}{t_{\theta}^2+1/2}, 
\label{eq_eom_r} \\
\theta'
= -\frac{1}{1-\beta}r^{\beta}\frac{t_{\theta}^2+3/2}{(t_{\theta}^2+1/2)(t_{\theta}^2+1)^{\beta/2}} - \frac{\alpha}{2} r^2 \frac{t_{\theta}}{t_{\theta}^2+1/2},
\label{eq_eom_theta}
\end{gather}
where $t_{\theta} \equiv \tan \theta$.
The Friedmann equation becomes
\begin{gather}
h^2
= \frac{1}{9}\mu\alpha (\chi^{'2} + \chi^n)
= \frac{1}{9}\mu\alpha (\chi_1^2 + \chi_2^2)
= \frac{1}{9}\mu\alpha r^2.
\label{eq_Friedmann_h}
\end{gather}
%%

%%%%%%%%%%%%%%%%%%%%%%%%%%%%%%%%%%%%%%%%%%%%%%%%%%
\subsection{Unperturbed solution}
%%%%%%%%%%%%%%%%%%%%%%%%%%%%%%%%%%%%%%%%%%%%%%%%%%

\begin{figure}
  \begin{minipage}{\columnwidth}
   \begin{center}
    \includegraphics[clip, width=0.5\columnwidth]{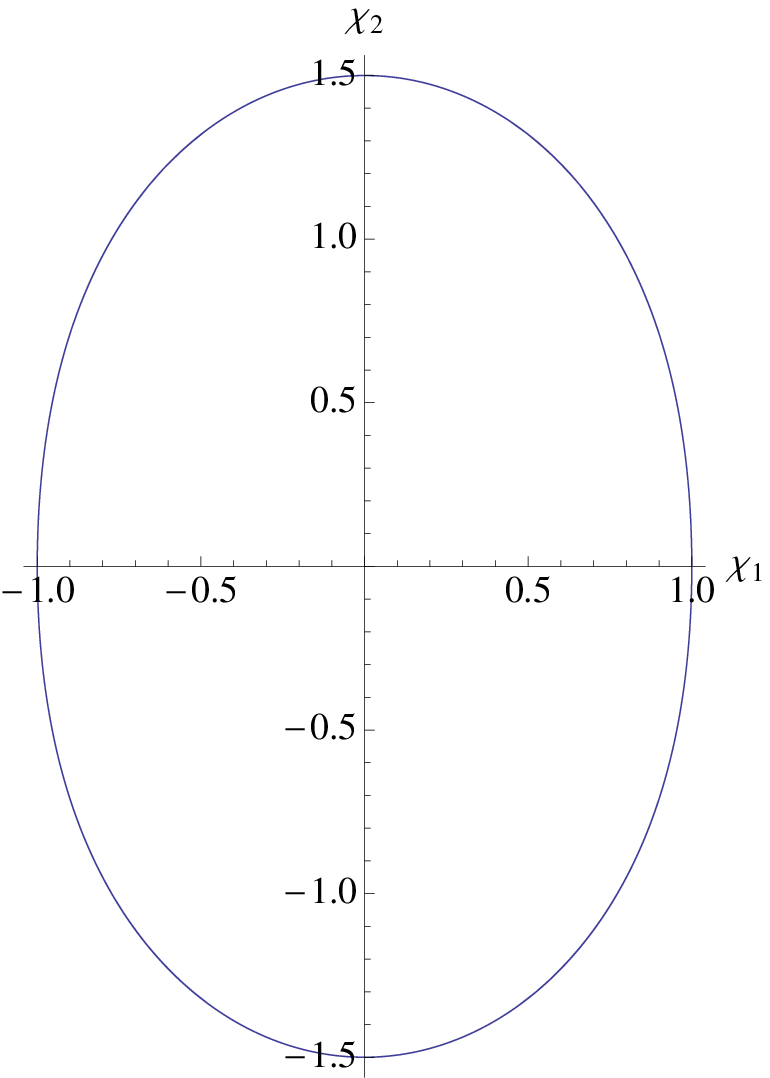}
   \end{center}
  \end{minipage}
  \caption{\small $\chi_1$-$\chi_2$ without perturbation.}
  \label{fig_chi12_unpert}
  \begin{minipage}{0.5\columnwidth}
    \begin{center}
      \includegraphics[clip, width=1.0\columnwidth]{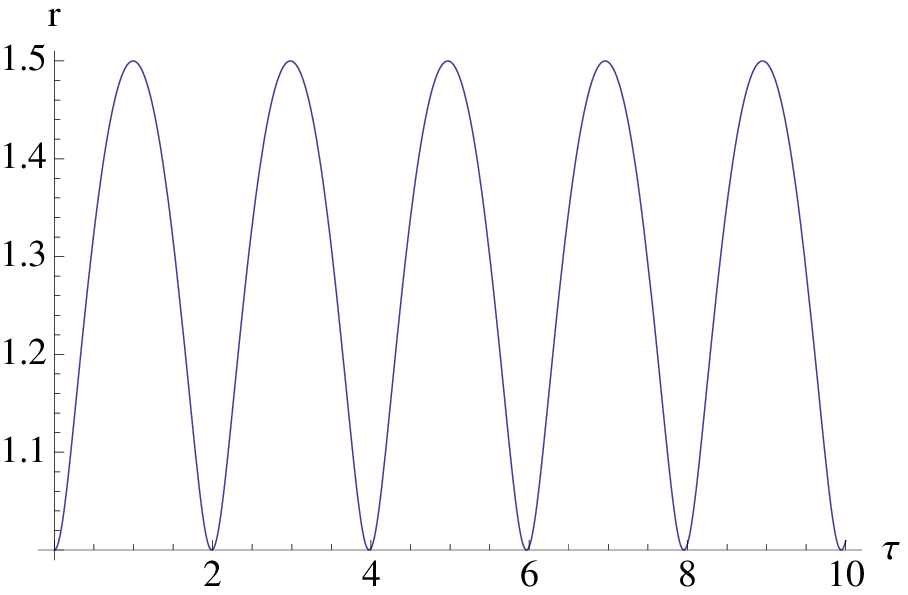}
    \end{center}
  \end{minipage}
  \begin{minipage}{0.5\columnwidth}
    \begin{center}
      \includegraphics[clip, width=1.0\columnwidth]{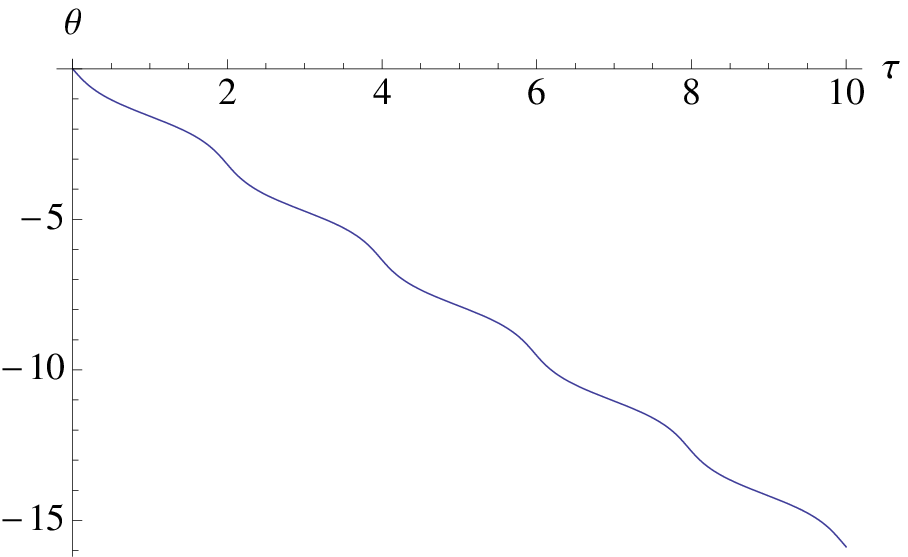}
    \end{center}
  \end{minipage}
  \caption{\small $\tau$-$r$ and $\tau$-$\theta$ without perturbation for $n=2$.}
  \label{fig_tau_rtheta_unpert_n=2}
\end{figure}

\begin{figure}
\begin{minipage}{0.5\columnwidth}
    \begin{center}
      \includegraphics[clip, width=1.0\columnwidth]{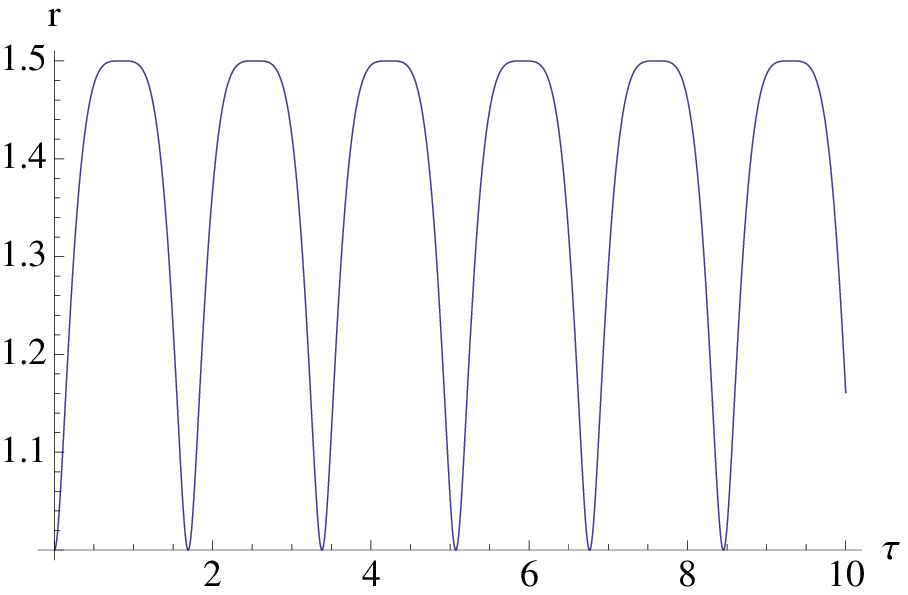}
    \end{center}
  \end{minipage}
  \begin{minipage}{0.5\columnwidth}
    \begin{center}
      \includegraphics[clip, width=1.0\columnwidth]{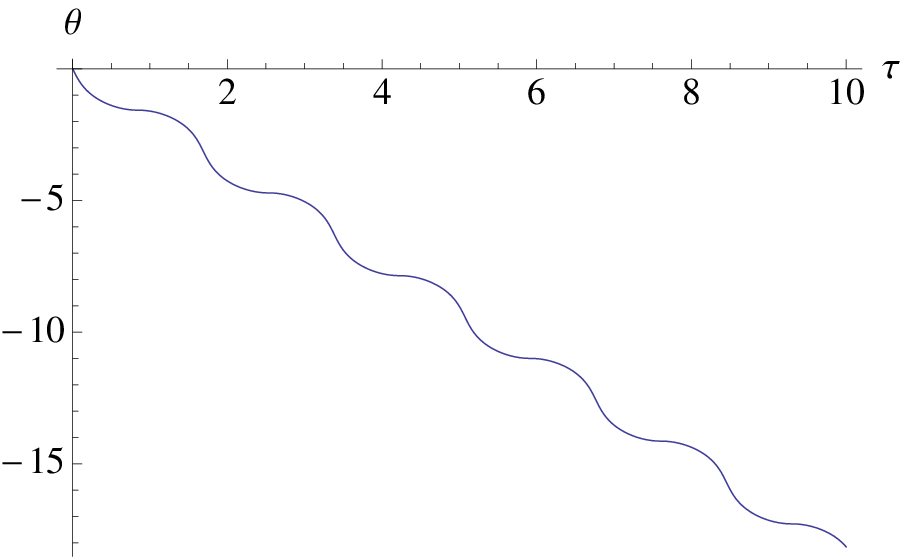}
    \end{center}
  \end{minipage}
  \caption{\small $\tau$-$r$ and $\tau$-$\theta$ without perturbation for $n=4$.}
  \label{fig_tau_rtheta_unpert_n=4}
\end{figure}

First, let us consider the solution of  Eqs.~(\ref{eq_eom_r}) and (\ref{eq_eom_theta}) in the limit of $\alpha = 0$. 
The relation between $r$ and $\theta$ is easy to obtain. Since
\begin{gather}
\frac{r'}{r}
= \frac{t_{\theta}}{(t_{\theta}^2+1)(t_{\theta}^2+3/2)}t_{\theta}',
\end{gather}
we obtain
\begin{gather}
r
= \frac{1+t_{\theta}^2}{1+2t_{\theta}^2/3}
= \left( \frac{5}{6} + \frac{1}{6}\cos 2 \theta \right)^{-1}.
\label{eq_unpert_r}
\end{gather}
The relation between $r$ and $\theta$ (or $\chi_1$ and $\chi_2$) is shown in Fig.~\ref{fig_chi12_unpert}.
Substituting Eq.~(\ref{eq_unpert_r}) into Eq.~(\ref{eq_eom_theta}), we get the relation between $\tau$ and $\theta$:
\begin{gather}
\tau
= -(1-\beta) \left[ 2 \gamma_{\tau} \left[ \frac{\theta}{\pi} + \frac{1}{2} \right]_{\rm F} + \delta_{\tau}(\theta) \right].
\label{eq_tau}
\end{gather}
Here $[\;]_{\rm F}$ is the floor function and
\begin{align}
\gamma_{\tau}
=& \lim_{t_{\theta} \rightarrow \infty} \delta_{\tau}(t_{\theta}), \\
\delta_{\tau} (\theta)
=& 2^{\beta} 3^{-\beta} \int_0^{t_{\theta}} dt_{\theta'} \frac{(t_{\theta'}^2+1/2)}{(t_{\theta'}^2+1)^{1+\beta/2}(t_{\theta'}^2+3/2)^{1-\beta}} \nonumber \\
=& t_{\theta} \left[ \frac{1}{3} {\rm F}_1 \left( \frac{1}{2},1+\frac{\beta}{2},1-\beta,\frac{3}{2},-t_{\theta}^2,-\frac{2t_{\theta}^2}{3} \right) \right. \nonumber \\
&\left.+\frac{2}{9} t_{\theta}^2{\rm F}_1 \left( \frac{3}{2},1+\frac{\beta}{2},1-\beta,\frac{5}{2},-t_{\theta}^2,-\frac{2t_{\theta}^2}{3} \right) \right], 
\end{align}
where F$_1$ is Appell's hypergeometric function. 
In $n=2$ case this relation reduces to 
\begin{eqnarray}
\gamma_{\tau}
&=& \left(-1+2\sqrt{\frac{2}{3}} \right) \frac{\pi}{2}, \\
\delta_{\tau} (\theta)
&=& -\arctan \left( \tan \theta \right) + 2\sqrt{\frac{2}{3}} \arctan \left( \sqrt{\frac{2}{3}} \tan \theta \right), \\
\tau
&=& \left(1-2\sqrt{\frac{2}{3}} \right) \theta 
+ \left[ \arctan \left( \tan \theta \right) - 2\sqrt{\frac{2}{3}} \arctan \left( \sqrt{\frac{2}{3}} \tan \theta \right) \right].
\label{eq_tau_theta_unpert}
\end{eqnarray}
Note that the relation between $\tau$ and $\theta$ is roughly
\begin{gather}
\tau
\simeq \left( 1 - 2\sqrt{\frac{2}{3}} \right) \theta
\end{gather}
in the limit of $t \rightarrow \infty$.
The relation between $\tau$ and $r$, and that between $\tau$ and $\theta$ for $n=2, 4$ are shown in Figs.~\ref{fig_tau_rtheta_unpert_n=2} and \ref{fig_tau_rtheta_unpert_n=4}.
We have checked that the numerical solution and the analytical one coincide with each other.

%%%%%%%%%%%%%%%%%%%%%%%%%%%%%%%%%%%%%%%%%%%%%%%%%%
\subsection{Perturbed solution}
%%%%%%%%%%%%%%%%%%%%%%%%%%%%%%%%%%%%%%%%%%%%%%%%%%

\begin{figure}
  \begin{minipage}{\columnwidth}
   \begin{center}
    \includegraphics[clip, width=0.5\columnwidth]{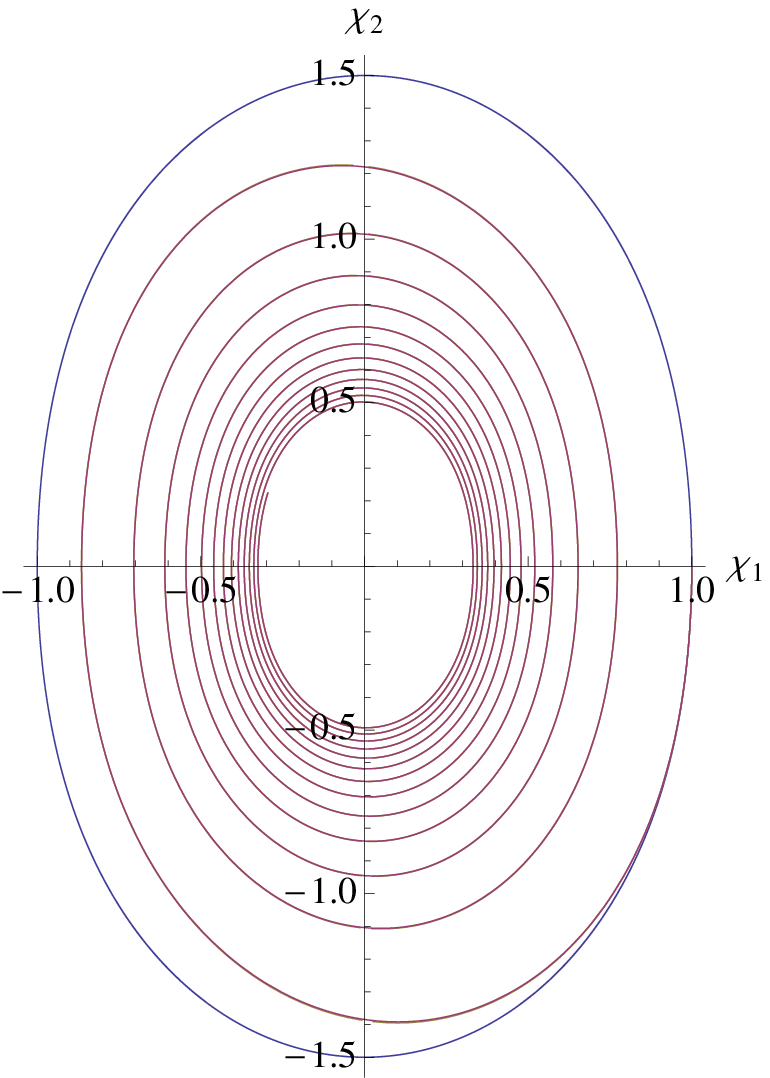}
   \end{center}
  \end{minipage}
  \caption{\small $\chi_1$-$\chi_2$ with perturbation for $n=2$ and $\alpha=0.15$. 
  (Blue) Unperturbed solution. (Red) Perturbed solution (numerical). (Yellow) Perturbed solution (analytical).
  Note that the red and yellow lines almost coincide with each other and we can hardly see the difference by eyes.}
  \label{fig_chi12_0.15_n=2}
  \begin{minipage}{0.5\columnwidth}
    \begin{center}
      \includegraphics[clip, width=1.0\columnwidth]{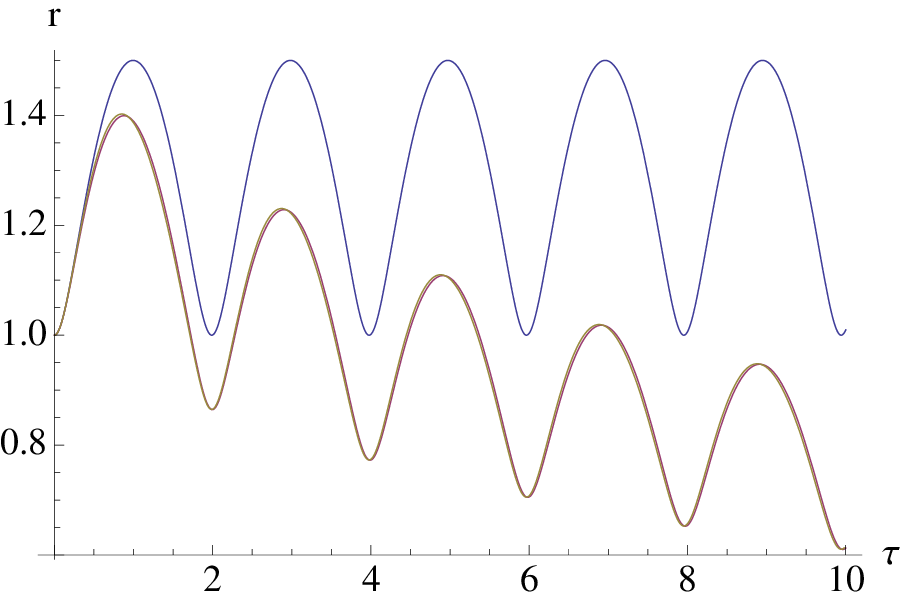}
    \end{center}
  \end{minipage}
  \begin{minipage}{0.5\columnwidth}
    \begin{center}
      \includegraphics[clip, width=1.0\columnwidth]{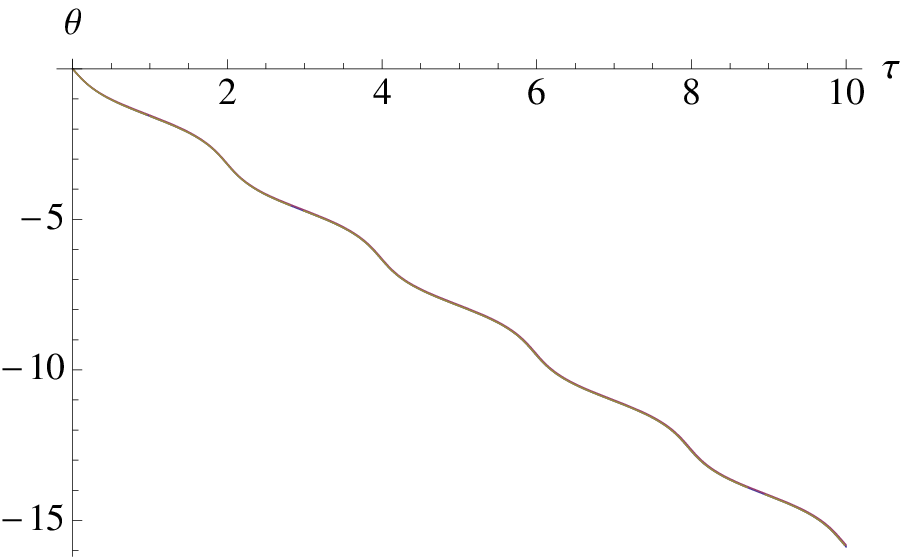}
    \end{center}
  \end{minipage}
  \caption{\small $\tau$-$r$ and $\tau$-$\theta$ with perturbation for $n=2$ and $\alpha=0.15$. 
  (Blue) Unperturbed solution. (Red) Perturbed solution (numerical). (Yellow) Perturbed solution (analytical).}
  \label{fig_tau_rtheta_0.15_n=2}
\end{figure}

\begin{figure}
   \begin{minipage}{0.5\columnwidth}
   \begin{center}
    \includegraphics[clip, width=1.0\columnwidth]{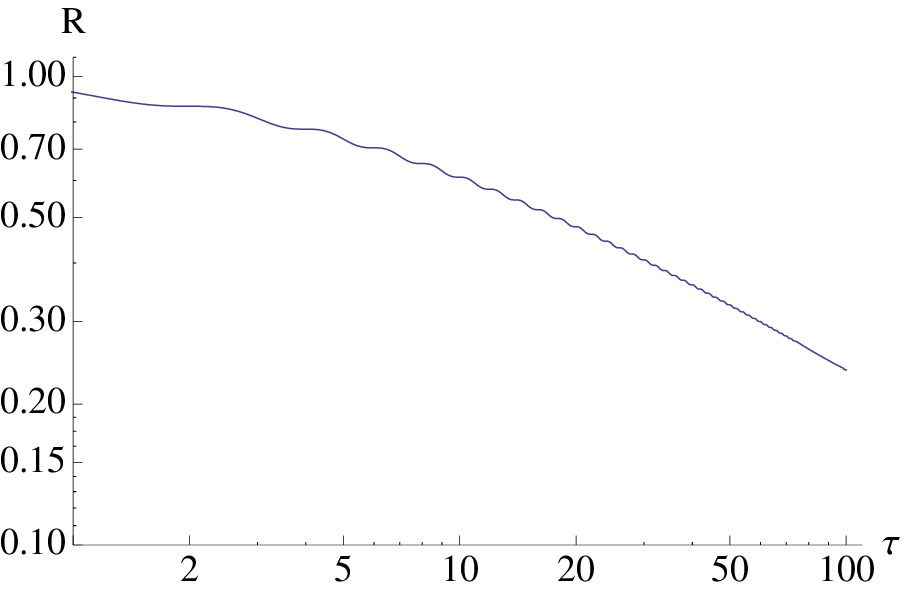}
   \end{center}
  \end{minipage}
     \begin{minipage}{0.5\columnwidth}
   \begin{center}
    \includegraphics[clip, width=1.0\columnwidth]{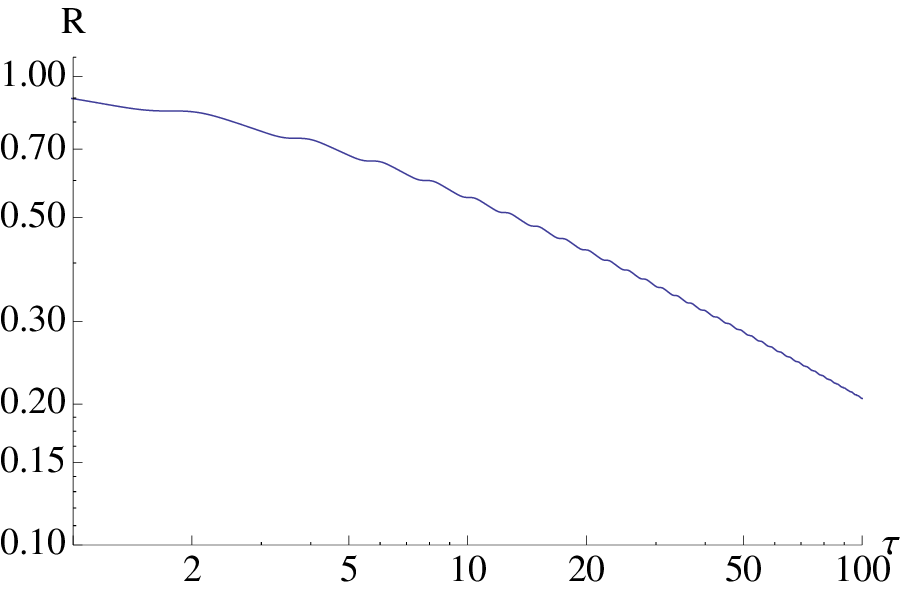}
   \end{center}
  \end{minipage}
  \caption{\small $\tau-R$ for $n=2$ (left) and $n=4$ (right). 
  Note that $R$ smoothly changes compared to $r$ and $\theta$ in the previous figures.}
  \label{fig_tau_capitalR_0.15}
\end{figure}

Next we include the $\alpha$-term in Eqs.~(\ref{eq_eom_r}) and (\ref{eq_eom_theta}). 
We substitute
\begin{gather}
r
= R(\theta) r_0(\theta)
= R(\theta) \frac{1+t_{\theta}^2}{1+2t_{\theta}^2/3},
\label{eq_r_Rtheta}
\end{gather}
where $r_0$ is the unperturbed solution, into these equations to obtain
\begin{eqnarray}
\frac{dR}{dt_{\theta}}
&=& \frac{\alpha(1-\beta)}{2} \left( \frac{3}{2} \right)^{2-\beta}
\frac{t_{\theta}^2(t_{\theta}^2+1/2)(t_{\theta}^2+1)^{1-\beta/2}}{(t_{\theta}^2+3/2)^{4-\beta}} R^3 \nonumber \\
&&\times \left[ R^{\beta}+\frac{\alpha(1-\beta)}{2}\left( \frac{3}{2} \right)^{2-\beta} 
\frac{t_{\theta}(t_{\theta}^2+1)^{2-\beta/2}}{(t_{\theta}^2+3/2)^{3-\beta}} R^2 \right]^{-1}.
\label{eq_eom_R}
\end{eqnarray}
Let us calculate the first-order solution to this equation.\footnote{
	In taking the limit $t \rightarrow \infty$ in the following, one might think that the
	term of $\mathcal O(\alpha^2)$ cannot be neglected.
	However, as one can see from terms inside the large parenthesis in Eq.~(\ref{eq_eom_R}),  
	the $\alpha$-term in the parenthesis becomes less and less effective for 
	$t \rightarrow \infty$ i.e. $R \rightarrow 0$. 
	In this sense the first order approximation is sufficient for the perturbative analysis,
	which is confirmed in the numerical calculation in the following.
}
Neglecting the $\alpha$-term in the large parenthesis, we can integrate the equation to get
\begin{gather}
\frac{1}{R^{2-\beta}}
= 1 - \alpha(1-\beta)(2-\beta) \left[2 \gamma_{\rm R} \left[ \frac{\theta}{\pi} + \frac{1}{2} \right]_{\rm F} + \delta_{\rm R} (\theta) \right], 
\label{eq_R}
\end{gather}
where
\begin{align}
\gamma_{\rm R}
=& \lim_{t_{\theta} \rightarrow \infty} \delta_{\rm R} (\theta), \\
\delta_{\rm R} (\theta)
=& 2^{-3+\beta} 3^{2-\beta} \int_0^{t_{\theta}} dt_{\theta'} 
\frac{t_{\theta'}^2 (t_{\theta'}^2+1/2)(t_{\theta'}^2+1)^{1-\beta/2}}{(t_{\theta'}^2+3/2)^{4-\beta}} \nonumber \\
=& t_{\theta}^3 \left[ \frac{1}{27} {\rm F}_1\left(\frac{3}{2},-1+\frac{\beta}{2},4-\beta,\frac{5}{2},-t_{\theta}^2,-\frac{2t_{\theta}^2}{3} \right) \right. \nonumber\\
&\left.+ \frac{2}{45} t_{\theta}^2 {\rm F}_1\left(\frac{5}{2},-1+\frac{\beta}{2},4-\beta,\frac{7}{2},-t_{\theta}^2,-\frac{2t_{\theta}^2}{3} \right) \right].
\end{align}
The relation between $R$ and $\theta$ for $n=2$ and $4$ is shown in Fig.~\ref{fig_tau_capitalR_0.15}. 
The relations between $\tau$ and $r$, and that between $\tau$ and $\theta$ are obtained after substituting Eqs.~(\ref{eq_r_Rtheta}) and (\ref{eq_R}) 
into Eqs.~(\ref{eq_eom_r}) and (\ref{eq_eom_theta}).
For $n=2$, the dominant first term in the RHS of Eq.~(\ref{eq_eom_theta}) does not have $r$ dependence. 
Therefore the relation between $\tau$ and $\theta$ remains almost unaltered compared with that for the unperturbed case.
Thus Eqs.~(\ref{eq_tau_theta_unpert}), (\ref{eq_r_Rtheta}) and (\ref{eq_R}) are a good analytical solution. 
One can confirm this fact in Fig.~\ref{fig_tau_rtheta_0.15_n=2}.

\begin{figure}
  \begin{minipage}{\columnwidth}
   \begin{center}
    \includegraphics[clip, width=0.5\columnwidth]{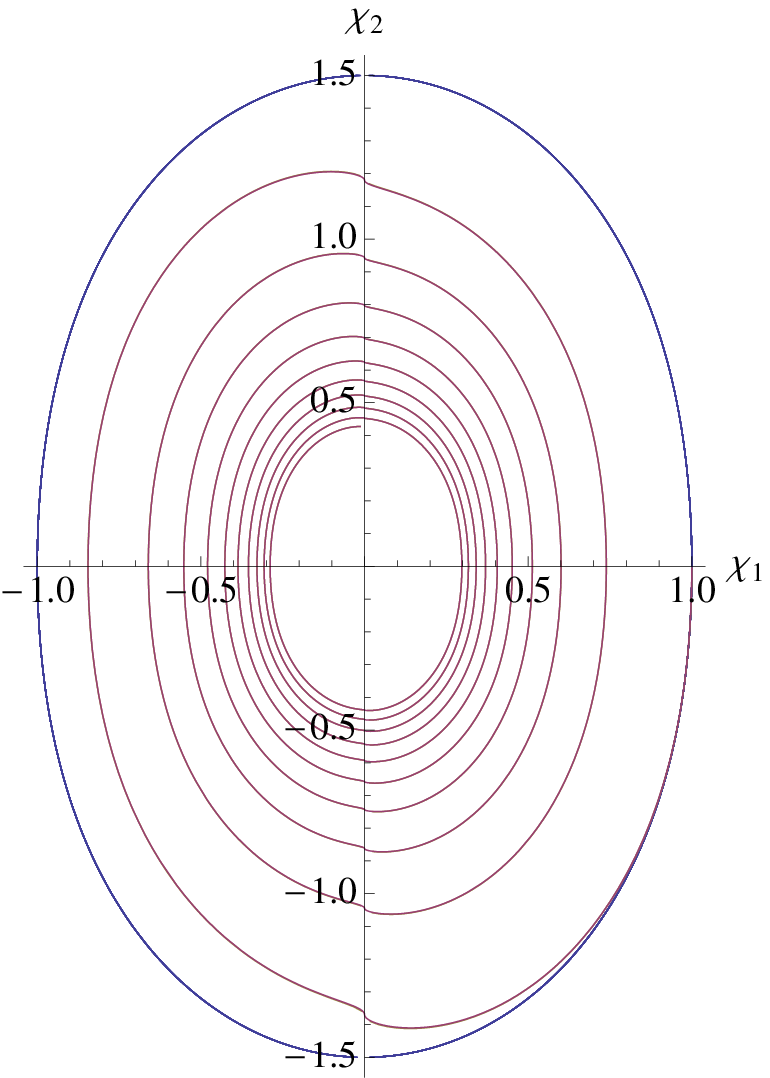}
   \end{center}
  \end{minipage}
  \caption{\small $\chi_1$-$\chi_2$ with perturbation for $n=4$ and $\alpha=0.15$. 
  (Blue) Unperturbed solution. (Red) Perturbed solution (numerical). (Yellow) Perturbed solution (analytical).
  Note that the red and yellow lines almost coincide with each other and we can hardly see the difference by eyes.}
  \label{fig_chi12_0.15_n=4}
  \begin{minipage}{0.5\columnwidth}
    \begin{center}
      \includegraphics[clip, width=1.0\columnwidth]{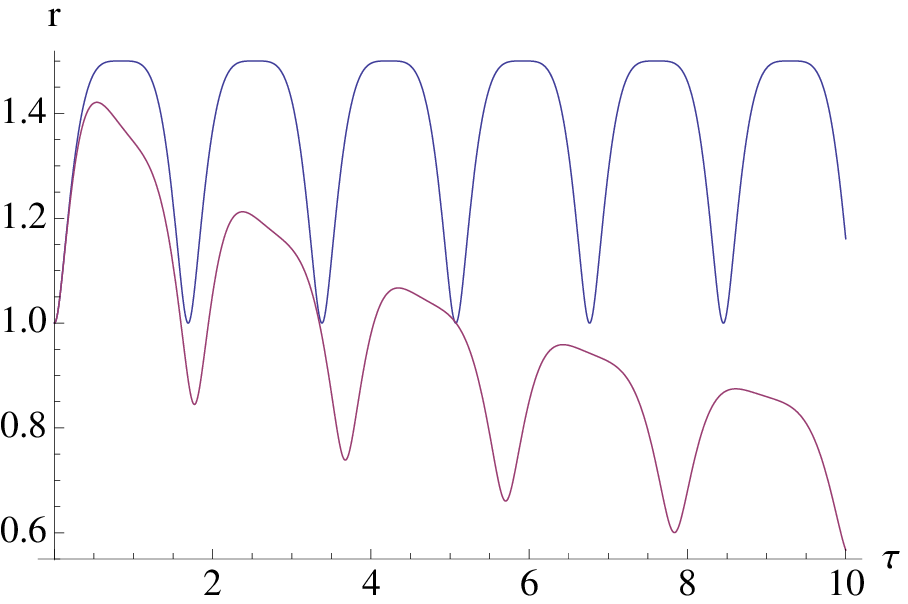}
    \end{center}
  \end{minipage}
  \begin{minipage}{0.5\columnwidth}
    \begin{center}
      \includegraphics[clip, width=1.0\columnwidth]{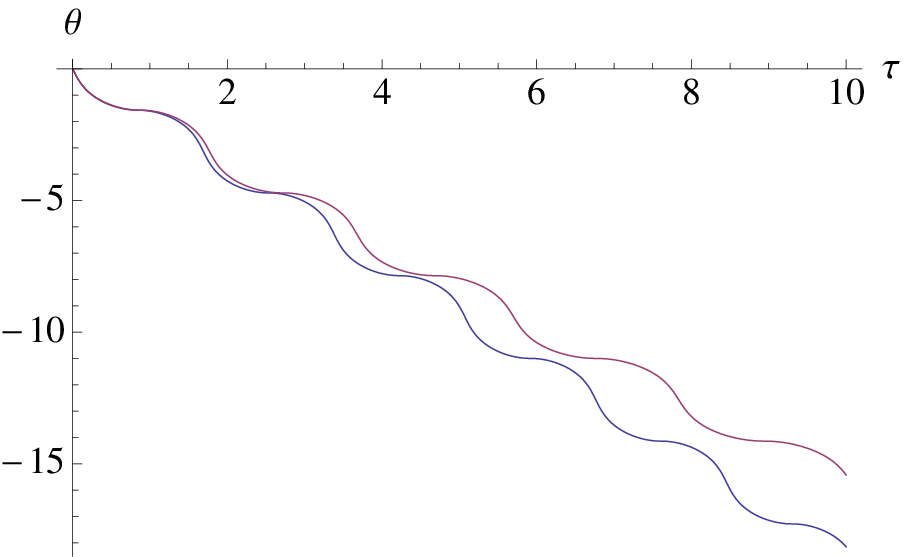}
    \end{center}
  \end{minipage}
  \caption{\small $\tau$-$r$ and $\tau$-$\theta$ with perturbation for $n=4$ and $\alpha=0.15$.  
  (Blue) Unperturbed solution. (Red) Perturbed solution (numerical).}
  \label{fig_tau_rtheta_0.15_n=4}
\end{figure}

As obvious from the discussion so far, 
\begin{gather}
R
= r \frac{1+2t_{\theta}^2/3}{1+t_{\theta}^2}
= r \left( \frac{5}{6} + \frac{1}{6} \cos 2 \theta \right)
= \frac{\chi_1^2+2\chi_2^2/3}{(\chi_1^2+\chi_2^2)^{1/2}}
= \frac{\chi^n+2\chi^{'2}/3}{(\chi^n+\chi^{'2})^{1/2}}
\end{gather}
is an adiabatic invariant, which is exacly conserved in the limit of $\alpha \to 0$
and changes slowly due to the non-vanishing $\alpha$ $(\sim H/m_{\rm eff})$.
One can easily show that $R$ is identical to $J$ in the main text, except for an overall constant.

%%%%%%%%%%%%%%%%%%%%%%%%%%%%%%%%%%%%%%%%%%%%%%%%%%
\subsection{Calculation of the coefficients $c,d$ and $e$}
%%%%%%%%%%%%%%%%%%%%%%%%%%%%%%%%%%%%%%%%%%%%%%%%%%

In this subsection we analytically calculate the coefficients $c,d$ and $e$ in the main text. 
The definitions Eqs.~(\ref{c})--(\ref{e}) read
\begin{eqnarray}
c
&=& \vev{\chi_1^2}, \\
d
&=& \frac{\vev{\chi_2^2}}{\vev{\chi_1^2}}, \\
e
&=& \frac{\vev{r^2}}{\vev{r}^2},
\end{eqnarray}
where $\vev{\;}$ means the $t$-average over a short period of oscillation\footnote{
	If one uses $\tau$-average instead, the results change slightly:
	for example, $\epsilon = 0.983$ and $0.822$ for $n=2$ and $4$, respectively. 
}
and the calculation can be done with the unperturbed solution.
The calculation is straightforward. For example,
\begin{eqnarray}
c
&=& \int d\theta \frac{d\tau}{d\theta} \frac{dt}{d\tau} \chi_1^2
\left/ \int d\theta \frac{d\tau}{d\theta} \frac{dt}{d\tau} \right. .
\end{eqnarray}
We substitute
\begin{eqnarray}
\chi_1
&=& r_0 \cos \theta, \\
\frac{d\tau}{d\theta}
&=& -(1-\beta)r_0^{-\beta} \frac{(t_{\theta}^2+1/2)(t_{\theta}^2+1)^{\beta/2}}{t_{\theta}^2+3/2}, \\
\frac{dt}{d\tau}
&=& \frac{(\mu \alpha)^{1/2}}{\mu M} r_0,
\end{eqnarray}
into the integral. The results are 
\begin{eqnarray}
c
&=& \left( \frac{3}{2} \right)^2 \frac{I\;[0,1,1-\beta/2,-4+\beta]}{I\;[0,1,-\beta/2,-2+\beta]}, \\
d
&=& \frac{I\;[1,1,1-\beta/2,-4+\beta]}{I\;[0,1,1-\beta/2,-4+\beta]}
= \frac{3n}{2}, \\
e
&=& \frac{I\;[0,1,-\beta/2,-2+\beta]\;I\;[0,1,2-\beta/2,-4+\beta]}{I\;[0,1,1-\beta/2,-3+\beta]^2},
\end{eqnarray}
where
\begin{gather}
I\;[\alpha,\beta,\gamma,\delta]
\equiv \int_0^{\infty} dt_{\theta} \;
(t_{\theta}^2)^{\alpha} (t_{\theta}^2+1/2)^{\beta} (t_{\theta}^2+1)^{\gamma} (t_{\theta}^2+3/2)^{\delta}.
\end{gather}
The last equality for $d$ is shown by noting that it can be reduced to an integral of a total derivative. 
The numerical values are summarized in Table~\ref{tab:gamma}.
We have checked that these values are reproduced by numerically solving the equation of motion (\ref{eqofm_full}).
Using $d$ obtained above, we obtain the expansion law (\ref{explaw}) with
\begin{gather}
\epsilon
\equiv \frac{3+2d}{3d}
= \frac{2n+2}{3n}.
\label{eq_epsilon_1}
\end{gather}
This final result on the expansion law relies on the procedure of Sec.~\ref{sec:ana}.
Below we see that the same result is derived by using our analytical solution.
It gives a strong consistency check of our results.

%%%%%%%%%%%%%%%%%%%%%%%%%%%%%%%%%%%%%%%%%%%%%%%%%%
\subsection{Expansion law of the universe}
%%%%%%%%%%%%%%%%%%%%%%%%%%%%%%%%%%%%%%%%%%%%%%%%%%

In this subsection we calculate the oscillation-averaged behavior of 
the Hubble parameter $\vev{h}=\epsilon / Mt$ using the analytical solution. 
This works as a consistency check for the previous subsection.

The Friedmann equation reads
\begin{gather}
\vev{h}
\simeq \frac{(\mu \alpha)^{1/2}}{3} \vev{R}\vev{r_0},
\label{eq_Friedmann_ave}
\end{gather}
where $r_0$ is given in Eq.~(\ref{eq_r_Rtheta}), and its oscillation average
\begin{gather}
\vev{r_0}
= \int d\theta \frac{d\tau}{d\theta}\frac{dt}{d\tau}r_0
\left/ \int d\theta \frac{d\tau}{d\theta}\frac{dt}{d\tau} \right.
\end{gather}
can be calculated as follows. 
First, by using
\begin{eqnarray}
\frac{d\tau}{d\theta}
&\simeq& -(1-\beta)\vev{R}^{-\beta}r_0^{-\beta}\frac{(t^2+1/2)(t^2+1)^{\beta/2}}{t^2+3/2}, 
\label{eq_dtaudtheta} \\
\frac{dt}{d\tau}
&\simeq& \frac{3}{\mu M} \vev{h}
\simeq \frac{(\mu \alpha)^{1/2}}{\mu M}\vev{R}r_0, 
\label{eq_dtdtau}
\end{eqnarray}
we can separate the rapidly oscillating part from the slowly oscillating one. 
Second, since we need the average in one oscillation period, we can neglect the 
time dependence of $\vev{R}$. 
Therefore we obtain
\begin{gather}
\vev{r_0}
\simeq \frac{3}{2} 
\int_0^{\infty} dt_{\theta} \frac{(t_{\theta}^2+1/2)(t_{\theta}^2+1)^{1-\beta/2}}{(t_{\theta}^2+3/2)^{3-\beta}} 
\left/ \int_0^{\infty} dt_{\theta} \frac{(t_{\theta}^2+1/2)(t_{\theta}^2+1)^{-\beta/2}}{(t_{\theta}^2+3/2)^{2-\beta}} \right. .
\end{gather}
The next task is to calculate the relation between $\vev{R}$ and $t$.
Substituting Eqs.~(\ref{eq_dtaudtheta}) and (\ref{eq_dtdtau}) into
\begin{gather}
t
= \int d\theta \frac{d\tau}{d\theta} \frac{dt}{d\tau},
\end{gather}
we get
\begin{eqnarray}
t
&\simeq& (1-\beta)\left( \frac{3}{2} \right)^{1-\beta} \frac{(\mu \alpha)^{1/2}}{\mu M} 
\int d\theta \frac{(t_{\theta}^2+1/2)(t_{\theta}^2+1)^{1-\beta/2}}{(t_{\theta}^2+3/2)^{2-\beta}} \vev{R}^{1-\beta} \nonumber \\
&\simeq& (1-\beta)\left( \frac{3}{2} \right)^{1-\beta} \frac{(\mu \alpha)^{1/2}}{\mu M} 
\frac{2}{\pi} \int_0^{\infty} dt_{\theta} \frac{t_{\theta}^2+1/2}{(t_{\theta}^2+1)^{\beta/2}(t_{\theta}^2+3/2)^{2-\beta}} 
\int_0^{\theta} d\theta \vev{R}^{1-\beta}.
\label{eq_t_Rave}
\end{eqnarray}
Note that $\vev{R}$ cannot be factored out in the first line since the integral is performed over many periods. 
Here we know from Eq.~(\ref{eq_R}) that
\begin{gather}
\vev{R}
\simeq \left[ \frac{2}{\pi} \alpha (1-\beta)(2-\beta) \gamma_{\rm R} \vev{-\theta} \right]^{-\frac{1}{2-\beta}},
\label{eq_Rave}
\end{gather}
in the limit of $t \rightarrow \infty$. 
Using Eqs.~(\ref{eq_t_Rave}) and (\ref{eq_Rave}) we get the relation between $\vev{R}$ and $t$, 
then substituting it into Eq.~(\ref{eq_Friedmann_ave}), we obtain\footnote{
	One can check the last equality by noting that it is reduced to the integral of a total derivative 
	as before.
}
\begin{eqnarray}
\vev{h}
&\simeq& \frac{\epsilon}{Mt}, \\
\epsilon
&\simeq& \frac{2}{3} \int_0^{\infty} dt_{\theta} \frac{(t_{\theta}^2+1/2)(t_{\theta}^2+1)^{1-\beta/2}}{(t_{\theta}^2+3/2)^{3-\beta}}
\left/ \int_0^{\infty} dt_{\theta} \frac{t_{\theta}^2(t_{\theta}^2+1/2)(t_{\theta}^2+1)^{1-\beta/2}}{(t_{\theta}^2+3/2)^{4-\beta}} \right. \nonumber \\
&=& \frac{2n+2}{3n},
\label{eq_epsilon_2}
\end{eqnarray}
the last of which is identical to Eq.~(\ref{eq_epsilon_1}).
This completes the consistency check of the expansion law given in Eq.~(\ref{explaw}).

%%%%%%%%%%%%%%%%%%%%%%%%%%%%%%%%%%%%%

%%%%%%%%%%%%%%%%%%%%%%%%%%%%%%%%%%%%%

\end{document}